\documentclass[twocolumn,showpacs,preprintnumbers,prl]{revtex4}
\usepackage {graphicx}

\begin{document}

\title {\bf Quasistatic Stick-slip in Dislocation Core and Frenkel-Kontorova Chain}

\author {M. Bhattacharya$^{\dagger\footnote{Email:mishreyee@vecc.gov.in}}$, A. Dutta$^\ddagger$,
and P. Barat$^{\dagger}$}

\affiliation {$^\dagger$Variable Energy Cyclotron Centre,1/AF Bidhannagar, Kolkata 700064, India \\
$^\ddagger$Department of Metallurgical and Materials Engineering,Jadavpur University, Kolkata 700032, India}

\date{\today}

\begin{abstract}

By means of atomistic simulations, we demonstrate that a dislocation core exhibits intermittent
quasistatic restructuring during incremental shear within the same Peierls valley. This can be regarded
as a \emph{stick-slip} transition, which is also reproduced for a one-dimensional Frenkel-Kontorova chain
under rigid boundary conditions. This occurs due to a discontinuous jump in an order parameter of the system,
which signifies the extent of region forbidden for the presence of particles in the chain. The \emph{stick-slip}
phenomenon observed in the dislocation core is also shown to be reflected after dimensionality reduction of
the multidimensional atomic coordinates, which provides a basis for comparison with the simple one-dimensional chain.

\end{abstract}


\maketitle

How much illustrative can we expect an ideal system to be, which is regarded as the simplified
model of a real physical phenomenon? The answer to this question is relevant in view of the
successful applications of such models throughout the history of physics. The Kronig Penney model of band structure,
Ising model for magnetic systems, liquid drop model of atomic nucleus, cellular automata model
of self-organized criticality etc. are some of the revolutionary models with remarkable achievements.
The Frenkel Kontorova (FK) chain \cite{frenkel, book} is one such celebrated model dealing with the
mechanics of discrete nonlinear systems. Conventionally, this model has successfully been employed
in studying a wide range of physical phenomena like the physics of dislocations \cite{ref1} and
crowdions \cite{ref2} in metals, adsorption of atoms on crystal surface \cite{ref3}, magnetic
structures \cite{ref4} etc. In addition, it has also rendered rich understanding of processes
like colloidal friction \cite{natmat2012} and biopolymers \cite{ref5, ref6} in recent years. \

The FK model consists of a linear chain of particles connected by springs and placed over
a substrate potential. In its most elementary form, the springs are assumed to be Hookean,
while a sinusoidal substrate potential is considered. It was first envisaged \cite{frenkel}
to represent the dislocation core as a kink in the FK chain. In particular, the existence of a
threshold force to move the kink directly corresponds to the Peierls stress \cite{peierls} for
the dislocation motion. The Peierls stress for a dislocation is often computed by means of
atomistic simulation \cite{caibook, sim1, Voskoboinikov}, where the incremental shear stress
(or strain) is applied quasistatically to the crystal at $T = 0$ K until the dislocation moves
to the next lattice site from the previous one. Similar quasistatic simulation of the
FK chain \cite {book,sim3} also yields the minimum force required to move the chain. Thus,
the FK model thematically depicts the mechanism of crossing the Peierls barrier in perspective
of discreteness and nonlinearity intrinsic to the lattice. Nevertheless, it is obscure whether
this simplified model is capable of providing finer details of the process. A closer look into
this issue is worthwhile as it would not only highlight the extent of resemblance between the
realistic physical system and the representative model, but also enable us to extend the
applicability of the model beyond a coarse representation. In this Letter, we study the atomistic
simulation of forcing a dislocation core out of its Peierls valley to obtain the atomic trajectories
at sub-Burgers vector resolution. Interestingly, the simulations reveal the occurrence of intermittent
relaxation bursts at such fine scale. Aiming to unfold the underlying mechanism, we apply the
appropriate boundary conditions to a simple one-dimensional FK chain to let it mimic the boundary
conditions used in the atomistic simulation. Even though the FK model is regarded
merely as a conceptual tool subsuming the inherent discreteness and nonlinearity of the lattice,
it also exhibits the abrupt bursts of structural relaxation, similar to the dislocation core.
In what follows, we shall argue that such a striking feature can be perceived as the quasistatic
counterpart of the stick-slip motion frequently observed in numerous dynamic systems
\cite{stickslip1, stickslip2, stickslip3, stickslip4}.
In addition, the technique of principal component analysis (PCA) \cite{pcabook} has been used
in an innovative way to establish the correspondence between a real physical system and its ideal model of lower
dimensionality.  \

The Peierls stresses, which are defined at absolute zero temperature, have been computed here
\cite{md++} for the four metals, molybdenum, iron, aluminium and copper. The simulation scheme
is akin to that used in the earlier measurements \cite{caibook, Voskoboinikov}. An edge dislocation
is introduced in a slab of finite thickness with periodic boundaries along the directions of dislocation
line and Burgers vector. The $x$, $y$ and $z$ dimensions of the b.c.c. simulation cells are
$90.5a\langle 111 \rangle /2$ , $40a\langle \bar{1}01 \rangle$  and $5a\langle 1\bar{2}1 \rangle$,
where $a$ is the lattice constant. The Burgers vectors and the dislocation lines are along the
$x$ and $z$ directions respectively [see the inset of Fig. 1(a)]. Corresponding cell dimensions for
the f.c.c. systems are
$90.5a\langle 101 \rangle /2$ $\times$ $20a\langle \bar{1}11 \rangle$ $\times$ $5a\langle 12\bar{1} \rangle$,
where the perfect dislocations splits into Schockley partials on account of the Frank's
criterion \cite{hirth}. Shear strain in the system is increased in small steps by tilting the
vertical boundary of the simulation cell. At each step, the system is relaxed to the minimum
energy configuration using the conjugate gradient method \cite{caibook}, while the top and bottom
surfaces are kept fixed during the relaxation. The potential energy of the system
is recorded and the crossover of the dislocation core to the next Peierls valley is marked
by a drastic drop in the total potential energy profile. Detailed analysis of the simulation
output involves the region only up to the crossover points, as shown in Fig. 1(a).
The dislocation core atoms are identified by using suitable centrosymmetric deviation
parameter (CSD) windows \cite{caibook}. For b.c.c Mo and Fe, the CSD window of range
1.4--10 {\AA}$^2$ has been used, while the ranges 3--20 {\AA}$^2$ and 3--16 {\AA}$^2$ are
employed for Al and Cu respectively \cite{com1}.\

\begin{figure}
\centerline{\includegraphics*[width=7cm, angle=0]{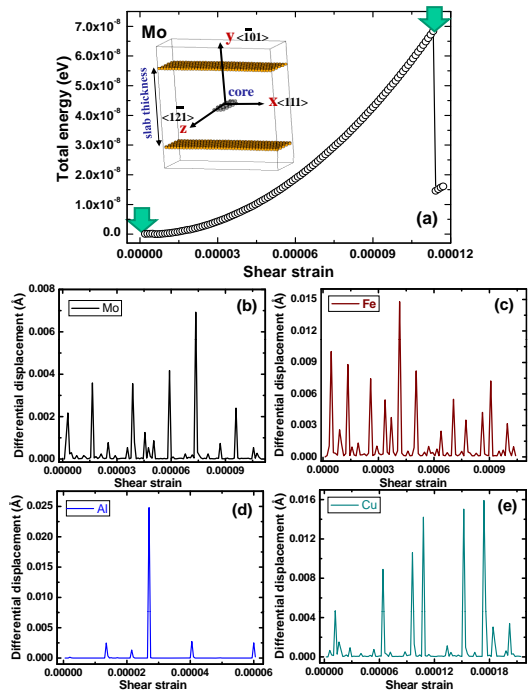}}
\caption{(color online). (a) Rise in potential energy of the Mo crystal with a typical simulation cell in the
inset. Note the sudden drop marking the instant of crossover to the adjacent Peierls valley. Differential
displacement profiles for the dislocation core atoms in (b) Mo, (c) Fe, (d) Al and (e) Cu. }
\end{figure}

As the process of applying incremental shear strain is followed by relaxation, the atomic structure tends to
reconfigure so that the potential energy of the system is minimized. The core structure can be specified
by a set of $n_c$ vectors, say $\{\textbf{r}_i\}$ ($i=1,2,..,n_c$), where $N_c$ is the number of atoms
in the dislocation core. We can now quantify the extent of aggregate core displacement at the $n^{th}$
step of incremental shear strain with respect to the previous step as
$\Delta = \sqrt{\sum_{i=1}^{N_c} |\textbf{r}_i(n)-\textbf{r}_i(n-1)|^2}$.
These differential displacements of the dislocation cores are plotted with the shear strains
in Figs. 1(b-e). Surprisingly, one can identify the intermittent relaxation bursts characterized by
narrow peaks in the profiles. Obviously, a small value of $\Delta$ means that
$\textbf{r}_i(n)\approx \textbf{r}_i(n-1)$ implying only a small structural change in the dislocation
core from the previous step of applied strain, whereas a large value is indicative of drastic structural
rearrangement. This proves that instead of exhibiting a continuous response to the incremental strain,
the dislocation core structure remains almost locked in between two successive bursts. This quasistatic
phenomenon is apparently analogous to the stick and slip states of dynamic variables in many
physical processes of interest. \

\begin{figure}
\centerline{\includegraphics*[width=7.5cm, angle=0]{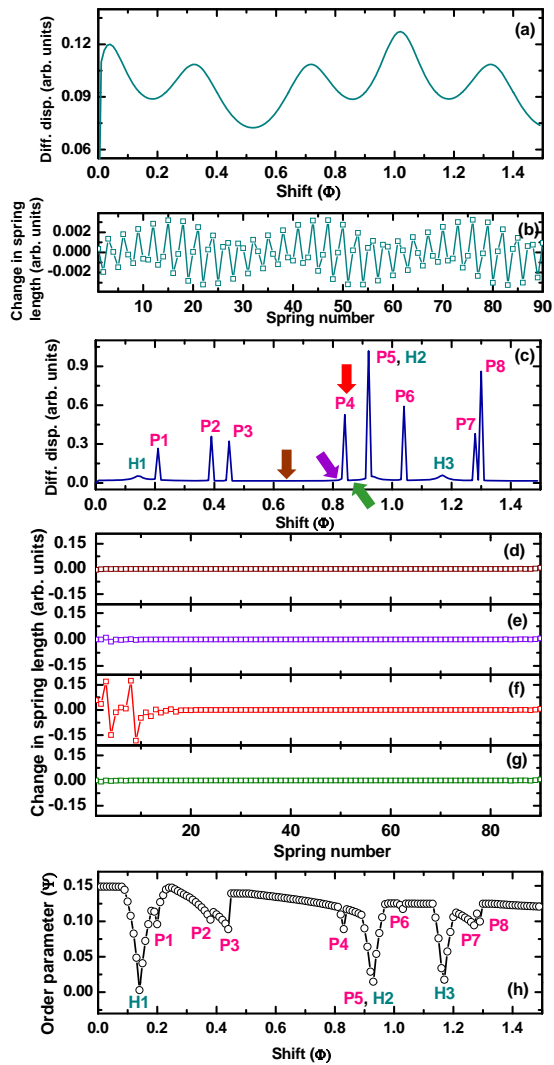}}
\caption{(color online). (a) Differential displacement and (b) change in the length of springs for
the FK chain accommodating $N$ = 91 particles in $N_v$ = 60 valleys with $E/\kappa$ = 0.01.
(c) At $E/\kappa$ = 0.035, the differential displacement profile shows intermittent peaks (P1-P8)
along with three small humps (H1, H2, H3). (d-g) Changes in the lengths of the 90 springs
showing the transition from the stick states (d,e) to the slip state
(f) at the fourth relaxation burst (P4), followed by another stick state (g) corresponding to the instants
indicated by the four arrows in (c). (h) Variation in the order parameter $\Psi$ as a function of shift
in the chain, $\Phi$. The features corresponding to the humps and peaks in (c) are marked here as well.
The peaks in (c) always coincide with discontinuous jumps in $\Psi$.  }
\end{figure}

In this study, the b.c.c. metals Mo and Fe are simulated using the modified Finnis-Sinclair interatomic
potential \cite{finnis}, whereas the glue potential \cite{glue} and an embedded atom model \cite{eam} are
used for Al and Cu respectively. In spite of a variety of potentials used in these computations, the relaxation
bursts can be observed in all the studied systems. This suggests that the origin of these bursts lies rather
in a more fundamental physical phenomenon, than in the complexities of the potential models employed
in the simulations. In this context, the possibility of closely investigating the FK model arises
as it reflects the basic features of a dislocation core, namely,  the nonlinear nature of interactions
and discreteness of the lattice. The potential energy of a finite FK chain with $N$ connected particles is
$U =  \sum_{i=1}^{N}\left[\frac{\kappa}{2}(x_{i+1}-x_{i}-l)^2 + E (1 - cos \frac {2\pi x_i}{b})\right]$,
where the first term denotes the harmonic spring potential with $x_i$, the position of $i^{th}$ particle,
$l$, the equilibrium length of each spring and $\kappa$, the spring constant. The second term
represents the periodic substrate potential where $b$ and $E$ are the periodicity and magnitude
of the potential respectively. Here we assume the natural length of the springs connecting the adjacent
atoms to be equal to the periodicity of the substrate potential (\emph{i.e.}, $l$ = $b$ = 1 arb. unit).
At this point, it should be noted that
despite its success in demonstrating the origin of the Peierls barrier, the FK chain does not exhibit a
direct correspondence to the dislocation core in all aspects. For instance, the long range interactions
between two dislocations \cite{hirth} is in clear contrast to the exponential short range kink-kink
interactions in the FK chain \cite{book}. Thus, a judicious choice of boundary conditions and parameters of the FK chain
is necessary so that the system can distinctly show the features of interest. In the atomistic simulations,
the top and bottom surfaces of the crystalline slabs were kept fixed during the relaxation process so that
the system could not revert back to the previous state of shear strain. To implement this on our FK chain,
we fix both of its ends, thereby yielding a fixed length and fixed density condition. We accommodate $N$
particles in $N_v$ valleys so that the coverage parameter $N/N_v$ is close to $\sim\frac{3}{2}$ to
approximately resemble the local coverage due to the extra half-plane of atoms at the core of an edge
dislocation. The chain is gradually shifted in small steps of $\delta$ and relaxed after each shift keeping
the first and the last particles rigid during the relaxation process. Hence, the net shift of the chain
at the $n^{th}$ step can be represented by the coordinates of the fixed end particle as $\Phi=x_1=n\delta$,
where the incremental shift ($\delta$) should have been infinitesimally small for the ideal quasistatic process.
However,
due to the trade-off between computational time and resolution, a small value of $\delta$ = 10$^{-2}$
(arb. units) is chosen and found to be sufficient to produce the requisite spatial resolution. Moreover,
in the present case of smooth periodic substrate potential, the simple
steepest-descent algorithm \cite{caibook} reasonably yields the relaxed states.\

The differential displacements are now computed for the particles of the FK chain for
different $E/\kappa$ ranging from 0.01 to 0.1. Fig. 2(a) shows the differential displacement for
$E/\kappa$ = 0.01 as a function of the shift, $\Phi$, imparted to the chain where a continuous
wavy nature can be observed. In addition, the change in
the lengths of the connecting springs in between two successive steps is shown in Fig. 2(b), where the
change is noticed along the entire chain. With increase in the value of $E/\kappa$, this wavy nature
changes to relaxation bursts characterized by the intermittent peaks, which can be seen
from Fig. 2(c) for $E/\kappa$ = 0.035. Remarkably, these peaks which demonstrate the
phenomenon of quasistatic stick-slip are present here also, similar to those in Figs. 1(b-e).
Corresponding changes in the length of springs, which are confined to the regions near the fixed ends
of the chain [Fig. 2(f)] are also intermittent and synchronized with the occurrence
of relaxation bursts.\

\begin{figure}
\centerline{\includegraphics*[width=6cm, angle=0]{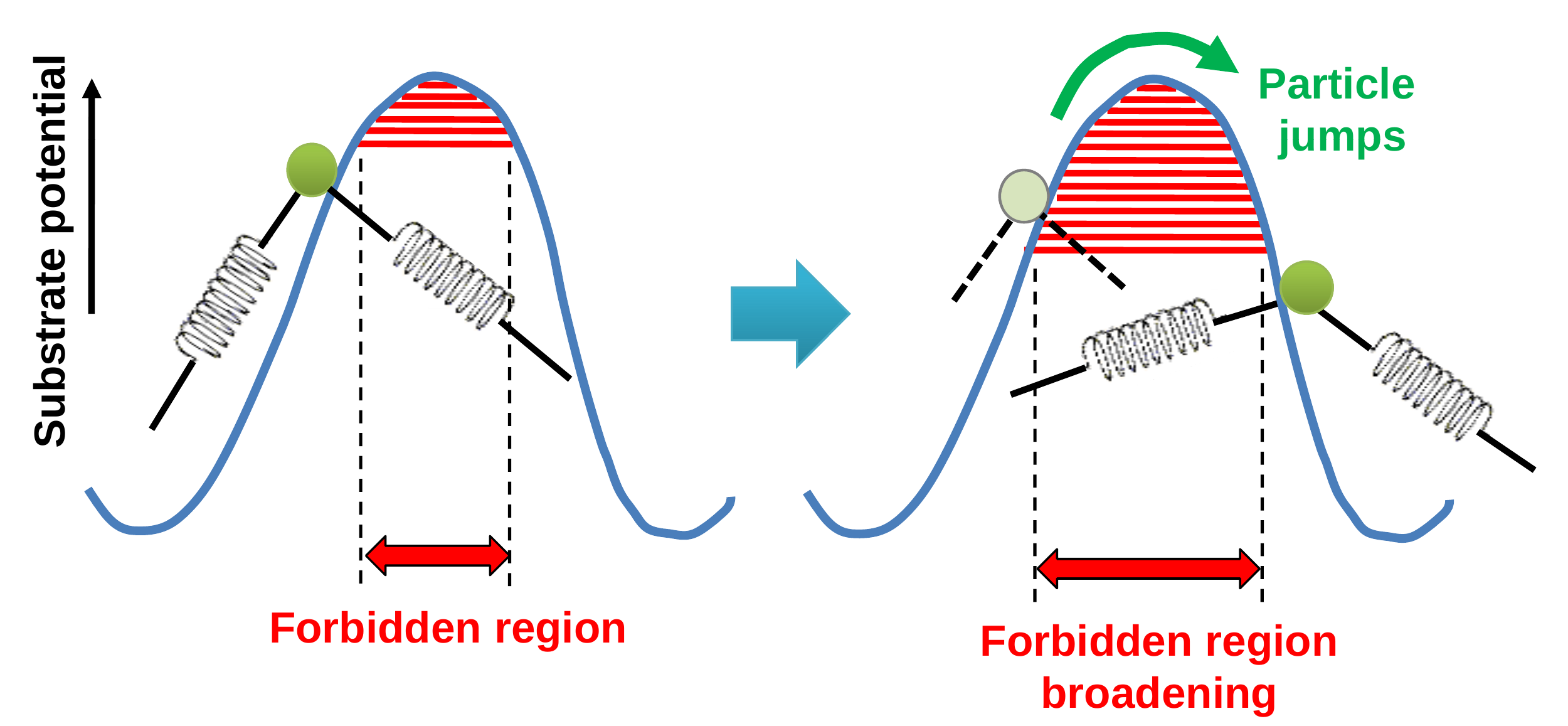}}
\caption{(color online). Schematic representation of discontinuous rise in the order parameter $\Psi$ as
the sudden broadening of the forbidden region. This causes the particle in the FK chain to abruptly cross the
potential hill.}
\end{figure}

For the FK chain in its ground state, the minimum distance of the particle from the nearest maximum of the
substrate potential has been used as an order parameter by Coppersmith and Fisher \cite{copper} to characterize
the Aubry-transition and breaking of analyticity \cite{book}. For the plot shown in Fig. 2(c), $E/\kappa$ is
large enough to cause the breaking of analyticity. However, in this study we encounter another variable
$\Phi$, denoting the extent of shift in the chain, which determines the ground state configuration
of the entire chain. Therefore, the order parameter given by $\Psi = min |x_i (mod~1) - 0.5|_{i\neq 1,N}$
has been computed and displayed in Fig. 2(h) as a function of $\Phi$. A finite non-zero value of $\Psi$
signifies the span of region around the top of the substrate potential where the presence of a particle
is forbidden. It is noticeable that despite the apparently sudden occurrences of relaxation bursts,
the instant of peak in Fig. 2(c) is always preceded by a gradual drop in $\Psi$ [Fig. 2(h)] during
the {\it stick} state, thereby indicating the gradual narrowing of the forbidden region.
Thereafter a discontinuous jump in $\Psi$ coincides with the transition to {\it slip} state at which
the relaxation burst occurs. This is indicative of abrupt broadening of the forbidden regions
and consequently, we expect at least one particle to suddenly cross over the peak of the substrate potential
(see the schematic in Fig. 3) thereby causing a steep rise in the differential displacement profile.
In addition, Fig. 2(h) also shows three instances where $\Psi$ drops and reaches values close to zero,
and then rises continuously. Because of this continuous change
in the order parameter, we observe small humps in the differential displacement profile which
is in sharp contrast to the abrupt peaks, where the rise in $\Psi$ is discontinuous. This can be
somewhat difficult to identify, for example, during the second hump (H2), where it coincides with
occurrence of fifth peak (P5) in the observed profile.  \

Typical ground state configurations of the FK chain is given in Fig. 4(a), where most of the particles
follow a structural pattern associated with a highly stable energy state and attempt to maintain it despite
the incremental change in $\Phi$. As a result, the local configurations near the two fixed ends is out of skew
with the rest of the chain and total energy of the system increases until the next relaxation burst. During this
{\it stick} state the differential displacement of particles are found to be negligibly small and hence, the
coordinates of the particles, $x_i (mod~1)$ after relaxation, when plotted with respect to the coordinates
before energy minimization show the linear behavior [Fig. 4(b)]. However, during a relaxation burst some of
the coordinates are knocked out of the straight line pattern, as shown in Fig. 4(c). As predicted earlier,
there is always one or more particles which cross the peak of the substrate potential corresponding to
$x_i (mod~1) = 0.5$ as marked in the figure.\

\begin{figure}
\centerline{\includegraphics*[width=8cm, angle=0]{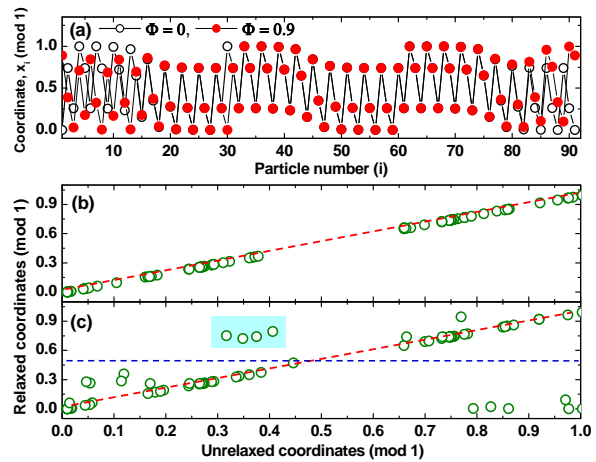}}
\caption{(color online). (a) Two typical ground state configurations of the FK chain ($\Phi$=0, 0.9) with $N$ = 91 particles
in $N_v$ = 60 valleys for $E/\kappa$ = 0.035. The relaxed coordinates of the particles (b) just before and
(c) at the instant of occurrence of the peak P5 in Fig. 2(c) plotted with respect to the unrelaxed coordinates.
The particles which have crossed the maxima of the substrate potential
i.e., $x_i (mod~1) = 0.5$ (shown by dotted horizontal line), are identified in the rectangular frame.  }
\end{figure}

The FK chain presented here is a 1-D model of an essentially 3-D atomic structure of a dislocation in crystal.
Each and every relaxation burst always drives the chain in a forward direction as both ends are shifted
in same direction. Similarly, it must be ascertained that the relaxation bursts as observed in the atomistic
simulation of dislocation core also drive the core in an effectively forward direction.
The notion of directionality is trivial in the
FK model because of the inherent single dimensionality associated with the structure. However, the motion
of a dislocation can be perceived only on a coarse scale of length, where the dislocation hops from one
lattice site to another. On a sub-Burgers vector scale of length, it is difficult to associate a sense of
{\it unidirectional} motion to the self-assembly of core atoms within the same Peierls valley unless a
specific directionality is ascribed to it. Clearly, the conventional technique of describing the core
position as the center of mass of all the core atoms (Refs. \cite{caibook, ourpaper1, ourpaper2} for example)
lacks the requisite resolution and alternative data mining tool needs to be explored. In this scenario,
we opt to use the PCA \cite{pcabook} as a prolific tool capable of providing a high degree of compressibility
of a high dimensional data and furnish its projection on a hyperspace of reduced effective dimensionality.
The versatility of the PCA is reflected in its successful applications across a wide range of studies
\cite{kim, wagner, hasbrouck, vaquila, kalinin} \emph{etc}. In the present case of
atomistic simulations, the coordinates of core atoms are recorded at each step of incremental shear
strain and this strain series data is arranged as a $n_s \times 3n_c$ matrix, where $n_s$ denotes
the number of strain steps and $n_c$ is the number of core atoms. Now each of the 3$n_c$ columns is
separately mean centered \cite{com3} and the mean-deviation matrix thus formed is used to generate
the covariance matrix \cite{pcabook}. Diagonalization of the covariance matrix yields the eigenvalues,
and the corresponding eigenvectors. Interestingly, for all the metals under study, the largest
normalized eigenvalues are always found to be in excess of 90$\%$. Such large values conclusively
prove a high compressibility intrinsic to these sets of multi-dimensional data. The projections of
the datasets along the principal directions corresponding to the largest eigenvalues are presented
in Fig. 5 for Mo and Al, for example. The sudden jumps present in these devil's staircase-like profiles are synchronized with
the occurrence of peaks in Fig. 1(b) and (d) and are typical signatures of the stick-slip process
(see the first and second panels of Fig. 4 in Ref. \cite{ref6}). Moreover, one can also observe
that each monotonic jump in the projected profile always causes a translation in
the same direction, thereby offering a ground for comparison with the FK model.   \

\begin{figure}
\centerline{\includegraphics*[width=8cm, angle=0]{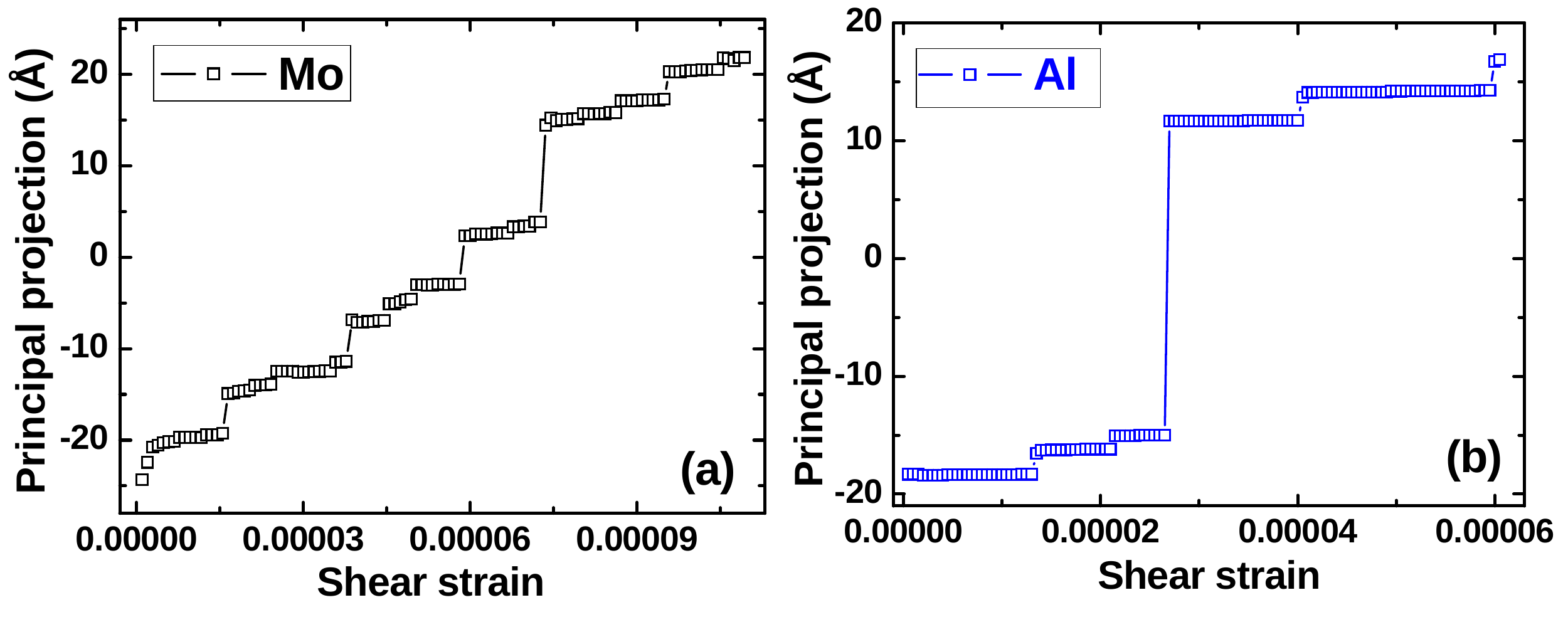}}
\caption{(color online). Principal projections for (a) molybdenum and (b) aluminium show staircase-like profiles.
The quasiplateaus represent the stick states whereas the slip states are reflected as the sudden jumps.  }
\end{figure}

To conclude, we have shown that at sub-Burgers vector resolution, intermittent relaxation bursts occur
in the quasistatic simulation of dislocation core. Similar features are observed for the
simple one-dimentional FK chain as well. This is attributed to a transition of the system from an effective {\it stick}
to {\it slip} state on account of abrupt broadening of the forbidden region around the peak of the substrate
potential. During the {\it stick-slip} transition, one or more atoms cross maximum of the substrate potential
and hops over the forbidden zone to cause a prominent rise in the differential displacement. Moreover, the tool of
principle component analysis has been used in an innovative way to extract the effective dimensionality
of the atomistic data of the dislocation core atoms. The projections of the atomic trajectories on the
principle directions further corroborate the efficacy of the 1-D FK chain in revealing the complex 3-D structure of the
dislocation core.
\\

A. Dutta acknowledges the financial support from CSIR, India to carry out this research work. \


\begin{thebibliography}{56}
\bibitem{com1} As there is no exact definition of the core region, the choice of the CSD window is
empirical. However, the results presented here are not very sensitive to the choice of filtering windows
so far as most of the core atoms are included.
\bibitem{com3} In the present study, we have actually made use of the so called `conditioned' data,
which means that the data is not only mean centered, but scaled so as to normalize the variances
as well. Although PCA would work only with mean centering also, complete conditioning is often
prescribed as a better way of preprocessing the data.
\end{thebibliography}
\end{document}